\documentclass[10pt]{article}
\usepackage{amssymb}
\usepackage{amsmath}
\usepackage{amsthm}
\usepackage{latexsym}
\usepackage[dvips]{epsfig}
\usepackage{mathrsfs}
\usepackage[bookmarks,colorlinks=false]{hyperref}

\theoremstyle{plain}
\newtheorem{proposition}{Proposition}
\newtheorem{remark}{Remark}
\newtheorem{lemma}{Lemma}
\newtheorem{theorem}{Theorem}

\newtheorem{definition}{Definition}

\newcommand{\updn}[3]{#1^{#2}_{\phantom{#2}#3}}

\begin{document}

\title{\textbf{Vacuum type D initial data}}

\author{{\Large Alfonso Garc\'{\i}a-Parrado G\'omez-Lobo} \thanks{E-mail address:
{\tt alfonso@math.uminho.pt}} \\
F\'{\i}sica Te\'orica, Universidad del Pa\'{\i}s Vasco,
Apartado 644, 48080 Bilbao, Spain,\\
Centro de Matem\'atica, Universidade do Minho,
4710-057 Braga, Portugal.
}
\maketitle

\begin{abstract}
A vacuum type D initial data set is a vacuum initial data set of the Einstein field equations whose data development contains a region 
where the space-time is of Petrov type D. 
In this paper we give a {\em systematic} characterisation of a vacuum type D initial data set.
By systematic we mean that the only quantities involved are those appearing in the vacuum constraints,
namely the first fundamental form (Riemannian metric) and the second fundamental form. Our characterisation is 
a set of conditions consisting of the vacuum constraints and some additional differential equations for the 
first and second fundamental forms. These conditions can be regarded as a system of partial differential equations 
on a Riemannian manifold and the solutions of the system contain all possible {\em regular} vacuum type D initial data 
sets. As an application we particularise our conditions for the case of vacuum data whose data development 
is a subset of the Kerr solution. This has applications in the formulation of the non-linear stability problem of the Kerr 
black hole.

\end{abstract}



\section{Introduction}
Vacuum type D solutions form a very important class of solutions of classical general relativity.
All the solutions of the family are known since long ago \cite{TYPE-D-KINNERSLEY}
and they have been studied extensively in the literature (see \cite{EXACTSOLUTIONS} for an account of the references). 
The most important member of the family is the Kerr solution \cite{KERR-METRIC} due to its 
physical interpretation as a spinning black hole. There is currently an ongoing effort to analyse the non-linear stability of 
the Kerr solution, a problem regarded as very hard which currently is wide open. The problem becomes somewhat easier if one 
studies particularisations thereof such as the analysis of axial perturbations of the 
Kerr solution, \cite{KLAINERMANAXIALKERR} or the linear stability of the Schwarzschild black hole
\cite{SCHWARZSCHILDLINEAR}.

Another line of attack could be to treat the nonlinear stability of the whole type D family, namely one considers perturbations
of data whose development is a vacuum solution of type D and then studies if the perturbed data development is close to a type D solution 
with no attention to which specific type D solution the development is close to. This approach involves dealing with a smaller number 
of algebraic and differential conditions as we only are imposing a restriction on the algebraic type of the Weyl tensor. Work along 
these lines can be consulted in \cite{TYPEDLINEAR, LINEARISEDD}.

In the present paper we give necessary and sufficient conditions for a vacuum initial data set to be {\em Petrov type D initial data}. What 
this means is that there exists an open set of the space-time containing the initial data hypersurface such that the Petrov type of the 
Weyl tensor is D (the precise definition is given in Definition \ref{def:typeD-initial-data}).
To find the conditions we start from a tensorial 
characterisation of Petrov type D vacuum solutions written in terms of a four rank tensor \cite{FERRANDOCOVARIANTPETROVTYPE} and 
project it down to the initial data hypersurface obtaining necessary conditions. To find the sufficient conditions we take advantage
of a property known to hold for type D space-times which is the existence of an invariant Killing vector field determined by the geometry.
Using the notion of {\em Killing initial data} we append to the necessary conditions additional conditions 
which guarantee that a Killing vector coinciding with the invariant Killing vector field exists in the data development. The 
property that the Lie derivative of any {\em Weyl concomitant} with respect to a Killing vector field vanishes enables us to 
propagate the necessary conditions and prove that the data development has an open subset of type D.
The characterisation of Petrov type D initial data obtained in this work only depends on the standard variables
used to set the vacuum initial value problem of the Einstein's equations, namely the Riemannian metric and 
the extrinsic curvature so we say that our characterisation is {\em algorithmic}. A characterisation of 
vacuum Petrov Type D data was found in \cite{GARKILLINGSPINOR} but it is written in the spinorial language requiring additional
variables and conditions in its formulation.

We stress that in the set-up of the non-linear stability problem of any vacuum solution of the 
Einstein's field equations it is necessary to find when the vacuum initial data are close to the data 
whose development yields the solution under study. For that we need an initial data characterisation
of the solution and the results presented in this work fulfill this goal for generic type D solutions 
and in particular for the case of the Kerr 
solution.

This paper is structured as follows: in section \ref{sec:typeD-characterisation} we review the tensor characterisation of vacuum
type D solutions presented in \cite{FERRANDOCOVARIANTPETROVTYPE} and obtain the expression of the complex invariant Killing vector
field in terms of the Weyl tensor (Theorem \ref{theorem:complex-killing}). The orthogonal splitting of these tensorial results is 
carried out in section \ref{sec:ot-typeD} which leads to a necessary set of conditions which any vacuum type D initial data has 
to satisfy (Theorem \ref{theo:type-d-initial-data-necessary}). The construction of necessary and sufficient conditions which
guarantee that the data development is of type D is carried out in section \ref{sec:typeD} and the result is 
presented in Theorem \ref{thm:type-d-initialdata-sufficient}. In section \ref{sec:applications} we  
find a local characterisation of the Kerr solution, presented in Theorem \ref{theo:kerr-local}, which enables us to compute 
additional restrictions to our main result about type D initial data in order to ensure that the data development is a subset of the 
Kerr space-time. All the algebraic tensor computations in this paper have been carried out with the {\em Mathematica} suite {\em xAct}
\cite{XACT}.

\section{Invariant chararacterisation of a vacuum type D space-time}
\label{sec:typeD-characterisation}
We shall work in a four-dimensional smooth manifold $(\mathcal{M},g_{\mu\nu})$, using Greek letters $\alpha,\beta,\dots$ to denote 
abstract tensor indices in the sense of Penrose. Our signature convention is $(-,+,+,+)$ and round brackets enclosing indices 
represent resp. anti-symmetrisation and symmetrisation. 
The Riemann and Ricci tensors are resp. $R^{\mu}{}_{\nu\alpha\beta}$, 
$R_{\mu\nu}$ and in this work the vacuum condition $R_{\mu\nu}=0$ will be assumed throughout. Hence the Riemann and the 
Weyl tensor $C_{\mu\nu\alpha\beta}$ coincide. We will use extensively 
complex valued tensor fields on the real manifold $\mathcal{M}$ and overbar will denote complex conjugation.

We introduce the self-dual Weyl tensor $\mathcal{C}_{\mu\nu\lambda\rho}$ which is given by 
\begin{equation}
\mathcal{C}_{\mu\nu\lambda\rho}\equiv
\frac{1}{2}(C_{\mu\nu\lambda\rho} - \mbox{i}\ C^*_{\mu\nu\lambda\rho}).
\label{eq:weyl-selfdual}
\end{equation}
We use the standard notation for the Hodge dual taken on a set of anti-symmetric indices of a tensor. The 
left and right dual of the Weyl tensor are the same and this leads to the 
self-duality property of $\mathcal{C}_{\mu\nu\lambda\rho}$
\begin{equation}
 \mathcal{C}^*_{\mu\nu\lambda\rho}= \ ^*\mathcal{C}_{\mu\nu\lambda\rho}=
 \mbox{i}\; \mathcal{C}_{\mu\nu\lambda\rho}
 \label{eq:self-duality}
\end{equation}
We introduce next the metric in the real vector space of self-dual Weyl-candidates (a Weyl candidate is any
tensor with the same algebraic properties as the Weyl tensor)
\begin{eqnarray}
 \mathcal{G}_{\mu\nu\rho\sigma}\equiv \frac{1}{2}(G_{\mu\nu\rho\sigma}-\mbox{i}\; \eta_{\mu\nu\rho\sigma})\;,
\end{eqnarray}
where  $G_{\mu\nu\rho\sigma}$ is given by 
\begin{equation}
 G_{\mu\nu\lambda\rho}\equiv g_{\mu\lambda} g_{\nu\rho} - g_{\mu\rho}g_{\nu\lambda}
\end{equation}

Our point of departure is the following result proven in \cite{FERRANDOCOVARIANTPETROVTYPE}

\begin{theorem}
A space-time is locally of Petrov type D if and only if
\begin{equation}
a\neq 0\;,\quad
\mathcal{D}_{\mu\nu\rho\sigma}\equiv \mathcal{C}_{\mu\nu\alpha\beta}\mathcal{C}^{\alpha\beta}_{\phantom{\alpha\beta}\rho\sigma}
-\frac{b}{a}\mathcal{C}_{\mu\nu\rho\sigma}-\frac{a}{6}\mathcal{G}_{\mu\nu\rho\sigma}=0\;,
\label{eq:define-D}
\end{equation}
where
\begin{equation}
 a\equiv \mathcal{C}_{\mu\nu\alpha\beta}\mathcal{C}^{\mu\nu\alpha\beta}\;,\quad
 b\equiv \mathcal{C}_{\mu\nu\alpha\beta}\mathcal{C}^{\alpha\beta}_{\phantom{\alpha\beta}\rho\sigma}
 \mathcal{C}^{\mu\nu\rho\sigma}
\end{equation}
\label{theorem:complex-type-D}
\end{theorem}

The tensor $\mathcal{D}_{\mu\nu\alpha\beta}$ has the following important algebraic property, 
which apparently has been overlooked in the literature.
\begin{proposition}
The tensor $\mathcal{D}_{\mu\nu\alpha\beta}$ is a self-dual Weyl candidate.
\label{prop:typeD-selfdual}
\end{proposition}
\proof We need to show that $\mathcal{D}_{\mu\nu\alpha\beta}$ has each of the algebraic properties which a self-dual
Weyl candidate should have.

\noindent
$\bullet$ It has the same mono-term symmetries as the Riemann tensor:
\begin{equation}
 \mathcal{D}_{[\mu\nu]\alpha\beta}=\mathcal{D}_{\mu\nu\alpha\beta}\;,\quad
 \mathcal{D}_{\mu\nu\alpha\beta}=\mathcal{D}_{\alpha\beta\mu\nu}.
\end{equation}
This is evident from (\ref{eq:define-D}) as each of the terms in the sum have these symmetries. 

\noindent
$\bullet$ It is traceless: one just needs to show that $\mathcal{D}^{\mu}{}_{\nu\mu\beta}=0$. This is achieved
by a direct computation combined with the {\em dimensionally dependent identity} which any Weyl candidate fulfills 
(see e. g. \cite{EDGARHOGLUNDDDI}, eq. (31))
\begin{equation}
\mathcal{C}_{\mu}{}^{\alpha\beta\gamma}\mathcal{C}_{\nu\alpha\beta\gamma}=\frac{1}{4}g_{\mu\nu}
\mathcal{C}^{\rho\alpha\beta\gamma}\mathcal{C}_{\rho\alpha\beta\gamma}.
\end{equation}

\noindent
$\bullet$
It satisfies the {\em cyclic property}: An explicit computation yields 
\begin{equation}
\mathcal{D}_{[\mu\nu\rho]\sigma}= 
\mathcal{C}_{\alpha\beta[\mu\nu}\mathcal{C}^{\alpha\beta}_{\phantom{\alpha\beta}\rho]\sigma}+\frac{\mbox{i}\ a}{12}\eta_{\mu\nu\rho\sigma}.
\label{eq:D-cyclic}
\end{equation}
On the other hand we have
\begin{equation}
\mathcal{C}_{\alpha\beta[\mu\nu}\mathcal{C}^{\alpha\beta}_{\phantom{\alpha\beta}\rho]\sigma}=
\mathcal{C}_{\alpha\beta[\mu\nu}\mathcal{C}^{\alpha\beta}_{\phantom{\alpha\beta}\rho\sigma]}.
\end{equation}
Any four-rank fully antisymmetric tensor must be proportional to the volume
element so 
\begin{equation}
 \mathcal{C}_{\alpha\beta[\mu\nu}\mathcal{C}^{\alpha\beta}_{\phantom{\alpha\beta}\rho\sigma]}=
 A \eta_{\mu\nu\rho\sigma}.
\end{equation}
The scalar factor $A$ is determined by contracting both sides with $\eta^{\mu\nu\rho\sigma}$ and 
using the self-duality of $\mathcal{C}_{\mu\nu\rho\sigma}$. One gets $A=-\mbox{i}\ a/12$ and from 
(\ref{eq:D-cyclic}) we conclude $\mathcal{D}_{[\mu\nu\rho]\sigma}=0$.

\noindent
$\bullet$
$\mathcal{D}_{\mu\nu\rho\sigma}$ is self-dual: this is also an explicit computation from 
eq. (\ref{eq:define-D}) which uses the self-duality of $\mathcal{C}_{\mu\nu\rho\sigma}$.
\begin{equation}
  \mathcal{D}^*_{\mu\nu\lambda\rho}= \ ^*\mathcal{D}_{\mu\nu\lambda\rho}=
 \mbox{i}\; \mathcal{D}_{\mu\nu\lambda\rho}.
\end{equation}
\qed

\begin{theorem}
If a vacuum space-time is of type D then there exists a complex vector field $\xi^\mu$ 
fulfilling the properties
\begin{equation}
\Xi {}_{\mu}{}_{\nu}= \frac{27}{2} w^{\frac{11}{3}} \xi {}_{\mu} \xi {}_{\nu}\;,
\quad \nabla_\mu\xi_\nu+\nabla_\nu\xi_\mu=0\;,
\label{eq:killing-type-D}
\end{equation}
where
\begin{eqnarray}
&&\Xi_{\mu\rho}\equiv \mathcal{Q}_{\mu\nu\rho\lambda}(\nabla^\nu w)(\nabla^\lambda w)\;,\quad
 \mathcal{Q}_{\mu\nu\rho\lambda}\equiv \mathcal{C}_{\mu\nu\rho\lambda}-w \mathcal{G}_{\mu\nu\rho\lambda}\;,
 \quad w\equiv-\frac{b}{2 a}.
 \label{eq:xiqw}
\end{eqnarray}

\label{theorem:complex-killing}
\end{theorem}
\proof If $\mathcal{C}_{\mu\nu\alpha\beta}$ has the algebraic type D then it can be written in the form 
\cite{FERRSAEZTYPED}
\begin{equation}
\mathcal{C}_{\mu\nu\alpha\beta}=6 w \mathcal{U}_{\mu\nu}\mathcal{U}_{\alpha\beta} + w\mathcal{G}_{\mu\nu\alpha\beta}\;,
\label{eq:weyl-canonical}
 \end{equation}
where $\mathcal{U}_{\mu\nu}$ is the so-called {\em canonical bi-vector} \cite{FERRSAEZTYPED}. For us its relevance 
is that under the Petrov type D condition it defines a complex Killing vector $\xi^\mu$ by the formula \cite{FERRSAEZTYPED} 
\begin{equation}
 \xi_\mu=w^{-\frac{1}{3}}\nabla_\nu\mathcal{U}^\nu_{\phantom{\nu}\mu}.
\label{eq:killingtobivector}
\end{equation}
We show next that the complex vector $\xi^\mu$ corresponds to the complex vector appearing in eq. (\ref{eq:killing-type-D}).
The canonical bi-vector is a complex self-dual 2-form and 
its real and imaginary parts are defined through the following equation
\begin{equation}
\mathcal{U}_{\mu\nu}=\frac{1}{\sqrt{2}}(U_{\mu\nu}-\mbox{i}\ U^*_{\mu\nu}).
\end{equation}
The real 2-form $U_{\mu\nu}$ is related to the Weyl tensor principal null directions $l_\mu$, $n_\mu$ by 
$U_{\mu\nu}=l_{[\mu}n_{\nu]}$ with a suitable normalisation for the null directions. 
Since in a Petrov type D spacetime both principal null directions are linearly independent
then we have the properties
\begin{equation}
 g_{\mu\nu} = 2 U_{\mu\alpha}U^{\alpha}_{\phantom{\alpha}\nu} - 2 (U^*)_{\mu\alpha}(U^*)^{\alpha}_{\phantom{\alpha}\nu}\;,\quad
 U_{\mu\alpha}(U^*)^{\alpha}_{\phantom{\alpha}\nu}=0.
\label{eq:Uproperties}
 \end{equation}
Now the second Bianchi identity $\nabla_\mu \mathcal{C}^\mu_{\phantom{\mu}\mu\rho\sigma}=0$ implies
\cite{FERRANDOCOVARIANTPETROVTYPE,FERRSAEZTYPED}
\begin{equation}
 6 \mathcal{U}_{\mu\nu}\nabla_{\rho}\mathcal{U}^{\rho\nu}=\nabla_\mu(\log(w)).
\end{equation}
We follow closely \cite{FERSAEZKERR} in the computations which follow.
Splitting the previous equation into real and imaginary parts we get
\begin{equation}
 U_{\mu\nu}\nabla_{\rho}U^{\rho\nu}-U^*_{\mu\nu}\nabla_{\rho}(U^*)^{\rho\nu}=R_\mu\;,\quad
 -U^*_{\mu\nu}\nabla_{\rho}U^{\rho\nu}-U_{\mu\nu}\nabla_{\rho}(U^*)^{\rho\nu}=\Theta_\mu\;,
\label{eq:obtainRTheta}
\end{equation}
where 
\begin{equation}
R_\mu\equiv \mbox{Re}\left(\frac{\nabla_\mu w}{3 w}\right)\;,\quad
\Theta_\mu \equiv \mbox{Im}\left(\frac{\nabla_\mu w}{3 w}\right).
\label{eq:defineRTheta}
\end{equation}
Combining (\ref{eq:obtainRTheta}) with (\ref{eq:Uproperties}) we obtain after some algebra
\begin{equation}
\nabla_\rho U^{\rho}{}_{\mu}=2 (U_{\mu\rho}R^\rho+U^*_{\mu\rho}\Theta^\rho)\;,\quad
\nabla_\rho (U^*)^{\rho}{}_{\mu}=2(U^*_{\mu\rho}R^\rho-U_{\mu\rho}\Theta^\rho).
\end{equation}
\begin{equation}
\nabla_\rho\mathcal{U}^{\rho}{}_{\mu}=2\mathcal{U}_{\mu\rho}(R^\rho+\mbox{i}\;\Theta^\rho)=
2\mathcal{U}_{\mu\rho}\frac{\nabla^\rho w}{3 w}.
\end{equation}
Using this last expression and the definition of the Killing vector $\xi_\mu$ we can compute the product 
$\xi_\mu\xi_\rho$ which results in
\begin{equation}
 \xi_\mu\xi_\rho=4 \mathcal{U}_{\mu\nu}\mathcal{U}_{\rho\lambda}\frac{\nabla^\nu w \nabla^\lambda w}{9 w^{\frac{8}{3}}}.
\end{equation}
Finally we replace here the product $\mathcal{U}_{\mu\nu}\mathcal{U}_{\rho\lambda}$ by 
its value found from (\ref{eq:weyl-canonical}) which yields (\ref{eq:killing-type-D}).
\qed

An interesting property of the Killing vector field which we shall require later on 
is \cite{FERRSAEZTYPED}
\begin{equation} 
[\vec{\boldsymbol\xi}\;,\overline{\vec{\boldsymbol\xi}}]=0. 
\label{eq:commute-killing}
\end{equation}

\section{Orthogonal splitting of the type D characterisation}
\label{sec:ot-typeD}
To fix the notation, we review first the standard notions used to compute 
the orthogonal splitting in general relativity (see e. g.  
\cite{ELLIS-COSMOLOGY,NATARIOCOSTAANALOGY,SE-DYNAMICALAWS} for full details).
Let $n^\mu$ be a unit time-like vector. 
The spatial metric is defined by $h_{\mu\nu}\equiv
g_{\mu\nu}+n_\mu n_\nu$ and it has the algebraic properties
$h^\mu_{\phantom{\mu}\mu}=3$,
$h_\mu^{\phantom{\mu}\sigma}h_{\sigma\nu}=h_{\mu\nu}$. We shall call a
covariant tensor $T_{\alpha_1\dots \alpha_m}$ {\em spatial} with
respect to $h_{\mu\nu}$ if it is invariant under
$h^\mu_{\phantom{\mu}\nu}$ i.e. if
\begin{equation}
h^{\alpha_1}_{\phantom{\alpha_1}\beta_1}\cdots h^{\alpha_m}_{\phantom{\alpha_m} \beta_m}T_{\alpha_1\cdots \alpha_m}=T_{\beta_1\cdots\beta_m}, 
\end{equation}
which is equivalent to the inner contraction of $n^\mu$ with $T_{\alpha_1\dots
\alpha_m}$ (taken on any index) vanishing. All this generalises straightforwardly 
for any mixed tensor. 
To find the orthogonal splitting of expressions containing
covariant derivatives we need to introduce the {\em spatial derivative} 
$D_\mu$ which is an operator whose action on any tensor
field $\updn{T}{\alpha_1\dots\alpha_p}{\beta_1\dots\beta_q}$,
$p,q\in\mathbb{N}$ is given by
\begin{equation}
D_\mu \updn{T}{\alpha_1\dots\alpha_p}{\beta_1\dots\beta_q}\equiv
h^{\alpha_1}_{\phantom{\alpha_1}\rho_1}\dots h^{\alpha_p}_{\phantom{\alpha_p} \rho_p}h^{\sigma_1}
_{\phantom{\sigma_1}\beta_1}\dots
h^{\sigma_q}_{\phantom{\sigma_q}\beta_q}h^{\lambda}_{\phantom{\lambda}\mu}
\nabla_\lambda
T^{\rho_1\dots\rho_p}_{\phantom{\rho_1\dots\rho_p}\sigma_1\dots\sigma_q}.
\label{spatialcd}
\end{equation}
From (\ref{spatialcd}) is clear that $D_\mu \updn{T}{\alpha_1\dots\alpha_p}{\beta_1\dots\beta_q}$ is spatial.

The orthogonal splitting
of a tensor expression consists in writing it as a sum of terms which
are tensor products of the unit normal and spatial tensors of lesser
degree or the same degree in which case the unit normal is absent. 
We write this statement for any tensor ${\boldsymbol T}$ as
\begin{equation}
{\boldsymbol T}=\sum_{J,P} {\boldsymbol T}^{(J)}_{(P)} {\boldsymbol n}_{J},
\label{eq:tensor-general-ot}
\end{equation}
where ${\boldsymbol n}_J$ represents a product of $J$-copies of the 1-form $n_\mu$ with appropriate abstract indices 
and ${\boldsymbol T}^{(J)}_{(P)}$ is a spatial covariant tensor with respect
to $n_\mu$. The index $P$ labels all possible spatial tensors appearing in the splitting.
If no factors $n_\mu$ are present then we set $J=0$. We recall the important property
\begin{equation}
{\boldsymbol T}=0\Longleftrightarrow
{\boldsymbol T}^{(J)}_{(P)}=0.
\label{eq:spatial-tensor-zero}
\end{equation}

We present next the orthogonal splitting of the most important tensorial quantities needed in this work.

\medskip
\noindent
$\bullet$ Orthogonal splitting of the volume element
\begin{equation}
\eta_{\alpha\beta\gamma\delta}=-n_\alpha\varepsilon_{\beta\gamma\delta}+n_\beta
\varepsilon_{\alpha\gamma\delta}
-n_\gamma\varepsilon_{\alpha\beta\delta}+n_\delta\varepsilon_{\alpha\beta\gamma}.
\label{decompose_eta}
\end{equation}
Here $\varepsilon_{\alpha\beta\gamma}\equiv n^\mu\eta_{\mu\alpha\beta\gamma}$ is the {\em spatial volume element} 
which is a fully antisymmetric spatial tensor.

\medskip
\noindent
$\bullet$ Orthogonal splitting of a {\em Weyl candidate}.
Any real or complex tensor,
$W_{\mu\nu\lambda\rho}$, with the same algebraic pro\-per\-ties as the Weyl tensor can be decomposed into its 
\emph{electric part}, $E(W)_{\mu\nu}$, and \emph{magnetic part},
$B(W)_{\mu\nu}$ as follows
\begin{equation}
W_{\mu\nu\lambda\sigma}=2\left(l_{\mu[\lambda} E(W)_{\sigma]\nu}-l_{\nu[\lambda}E(W)_{\sigma]\mu}
-n_{[\lambda} B(W)_{\sigma]\tau}\varepsilon^\tau_{\phantom{\tau}\mu\nu}
-n_{[\mu} B(W)_{\nu]\tau}\varepsilon^\tau_{\phantom{\tau}\lambda\sigma} 
\right)\;,\label{weyl-1generalized} 
\end{equation}
where
\begin{equation}
E(W)_{\tau\sigma}\equiv W_{\tau\nu\sigma\lambda}n^\nu n^\lambda, \quad 
B(W)_{\tau\sigma}\equiv W^*_{\tau\nu\sigma\lambda}n^\nu n^\lambda, \quad 
\label{eq:e-m-generic}
\end{equation}
and $l_{\mu\nu}\equiv h_{\mu\nu}+n_{\mu}n_{\nu}$.
The spatial tensors $E(W)_{\mu\nu}$ and $B(W)_{\mu\nu}$
are symmetric and traceless.

In the particular case of the self-dual Weyl tensor we have the definitions
\begin{equation}
\mathcal{E}_{\mu\nu} \equiv E(\mathcal C)_{\mu\nu}, \quad \mathcal{B}_{\mu\nu} \equiv B({\mathcal C})_{\mu\nu}.
\label{eq:weyl-electric-magnetic}
\end{equation}
These are not independent, given the self-duality property (\ref{eq:self-duality})
\begin{equation}
\mathcal{B}_{\mu\nu}=\mbox{i}\; \mathcal{E}_{\mu\nu}\;, 
\end{equation}
Therefore the splitting (\ref{weyl-1generalized}) in the case of the 
self-dual Weyl tensor becomes
\begin{equation}
\mathcal{C}_{\mu\nu\lambda\sigma}=2\left(l_{\mu[\lambda} \mathcal{E}_{\sigma]\nu}-l_{\nu[\lambda}\mathcal{E}_{\sigma]\mu}
-\mbox{i}\ n_{[\lambda}\mathcal{E}_{\sigma]\tau}\varepsilon^\tau_{\phantom{\tau}\mu\nu}
-\mbox{i}\ n_{[\mu} \mathcal{E}_{\nu]\tau}\varepsilon^\tau_{\phantom{\tau}\lambda\sigma} 
\right).
\label{eq:decompose-weyl-selfdual}
\end{equation}
This formula can indeed be used for any other self-dual Weyl candidate with the appropriate changes.
The tensor $\mathcal{E}_{\mu\nu}$ is related to the standard Weyl tensor 
electric, $E_{\mu\nu}$, and magnetic, $B_{\mu\nu}$, parts through the relation
\begin{equation}
 \mathcal{E}_{\mu\nu}=\frac{1}{2}(E_{\mu\nu}-\mbox{i}\; B_{\mu\nu})\;,
\label{eq:eselfdualtoenormal}
\end{equation}
which is a direct consequence of (\ref{eq:weyl-selfdual}). 
With the aid of (\ref{eq:decompose-weyl-selfdual}) any 
{\em algebraic concomitant} of the self-dual Weyl tensor can be rendered in terms of $\mathcal{E}_{\mu\nu}$. 
For example one has
\begin{equation}
a = 16 \mathcal{E}_{\mu\nu} \mathcal{E}^{\mu\nu}\;,\quad
b = -64 \mathcal{E}_{\mu}{}^{\alpha} \mathcal{E}^{\mu\nu} \mathcal{E}_{\nu\alpha}.
\label{eq:absplit}
\end{equation}

\medskip
\noindent
$\bullet$ Orthogonal splitting of the Bianchi identity $\nabla_{[\mu}\mathcal{C}_{\nu\rho]\sigma\beta}=0$.
We assume now that the {\em Frobenius condition} $n_{[\mu}\nabla_{\nu}n_{\sigma]}=0$ holds and 
define the spatial tensor
\begin{equation}
K_{\mu\nu}\equiv-\frac{1}{2}\mathcal{L}_{\vec{n}}h_{\mu\nu}\;,\quad
K_{(\mu\nu)} = K_{\mu\nu}.
\label{lie-h}
\end{equation}
Combining the previous definition with the Frobenius condition we get
\begin{equation}
\nabla_\mu n_\nu=-K_{\mu\nu}-n_\mu A_\nu,
\label{gradnormaltoextrinsic}
\end{equation}
where $A^\mu\equiv n^\rho\nabla_\rho n^\mu$ is the acceleration of $n^\mu$ and it is spatial. 
Using the orthogonal splitting (\ref{eq:decompose-weyl-selfdual}) and eq. (\ref{gradnormaltoextrinsic})  
we find that the Bianchi identity is equivalent to 
\begin{eqnarray}
&&\pounds_{\vec{\boldsymbol n}} \mathcal{E}_{\mu\nu} = 
 2 K^{\beta}{}_{\beta} \mathcal{E}_{\mu\nu} -2 i A^{\beta} \mathcal{E}_{(\mu}{}^{\delta}\varepsilon_{\nu)\beta\delta} - 
i \varepsilon_{(\mu}{}^{\beta\delta}D_{|\beta|}\mathcal{E}_{\nu)\delta} + 
\frac{1}{2} h_{\mu\nu} (K^{\beta\delta} \mathcal{E}_{\beta\delta} + K_{\beta}{}^{\delta} \mathcal{E}^{\beta}{}_{\delta})\nonumber\\
&&-5 K_{(\mu}{}^{\beta}\mathcal{E}_{\nu)\beta}\;,\label{eq:lied-weyl-electric}\\
&& D_{\delta}\mathcal{E}_{\alpha}{}^{\delta} = i \varepsilon_{\alpha\mu\beta} K^{\delta\mu} \mathcal{E}_{\delta}{}^{\beta}.
\end{eqnarray}

\begin{proposition}
 One has the equivalence 
 \begin{equation}
  \mathcal{D}_{\mu\nu\rho\alpha}=0 \Longleftrightarrow \mathfrak{a}_{\mu\nu}=0\;,\quad
\mathfrak{a}_{\mu\nu} \equiv \frac{a}{12} h_{\mu\nu} -  \frac{b}{a}\mathcal{E}_{\mu\nu} - 4 \mathcal{E}_{\mu}{}^{\alpha} 
\mathcal{E}_{\nu\alpha}.  
 \end{equation}
\qed
\label{prop:d-tensor-split}
\end{proposition}
\proof 
The tensor $\mathcal{D}_{\mu\nu\alpha\beta}$ is a self-dual Weyl candidate
as proven in Proposition \ref{prop:typeD-selfdual}. 
Hence we can use eq. (\ref{eq:decompose-weyl-selfdual}) to find its orthogonal splitting
with $\mathcal{E}_{\mu\nu}$ replaced by the following spatial tensor
\begin{equation}
\mathcal{D}_{\mu\alpha\nu\beta} n^\alpha n^\beta. 
\end{equation}
A computation using (\ref{eq:define-D}), (\ref{eq:decompose-weyl-selfdual}) gives
\begin{equation}
\mathcal{D}_{\mu\alpha\nu\beta} n^\alpha n^\beta=\mathfrak{a}_{\mu\nu}.
\end{equation}
The proposition is now a consequence of (\ref{eq:spatial-tensor-zero}).
\qed

\begin{proposition}
One has the decomposition
\begin{eqnarray}
&&\nabla_\mu w = n_\mu \omega^\parallel + \omega_\mu^\bot\;,\label{eq:decomposew}\\
&&(\omega^\bot)_\mu\equiv 
\frac{b D_ {\mu}\mathit{a} -  \mathit{a} D_ {\mu}\mathit{b}}{2a^2}=D_\mu w\;,\\
&&\omega^\parallel\equiv 
\frac{6 K^{\beta\gamma}}{\mathit{a}^3} \bigl(\mathit{b}^2 \mathcal{E}_{\beta\gamma} + \mathit{a} \
(\mathit{b} \mathfrak{a}_{\beta\gamma} - 12 \mathit{a} \mathcal{E}_{\beta}{}^{\rho} \
\mathfrak{a}_{\gamma\rho})\bigr) -w K^{\beta}{}_{\beta} + 16\mbox{i}\ 
\frac{(\mathit{b} \mathcal{E}^{\beta\gamma} + 3 \mathit{a} \mathfrak{a}^{\beta\gamma})}{2 \mathit{a}^2}
\varepsilon_{\gamma \rho\lambda} D^{\lambda}\mathcal{E}_{\beta}{}^{\rho}.\nonumber\\
&& 
\end{eqnarray}
\label{prop:omega-decomposition}
\end{proposition}
\proof
We start from the definition of $w$, eq. (\ref{eq:xiqw}), and replace $a$, $b$ by the expressions
shown in (\ref{eq:absplit}) getting
\begin{equation}
w=\frac{2 \mathcal{E}_{\mu}{}^{\alpha} \mathcal{E}^{\mu\nu} \mathcal{E}_{\nu\alpha}}{\mathcal{E}_{\mu\nu} \mathcal{E}^{\mu\nu}}.
\label{eq:w-split}
\end{equation}
We use this formula to compute the covariant derivative of $w$. The covariant derivative terms of $\mathcal{E}_{\mu\nu}$
are worked out with the formula
\begin{equation}
\nabla_{\alpha}\mathcal{E}_{\beta\gamma} = - 2 K_{\alpha}{}^{\delta} n_{(\gamma} \mathcal{E}_{\beta)\delta} 
- 2 n_{\alpha} A^{\delta} n_{(\gamma} \mathcal{E}_{\beta)\delta} + D_{\alpha}\mathcal{E}_{\beta\gamma} -
2n_{\alpha} \left(K_{(\gamma}{}^{\delta} \mathcal{E}_{\beta)\delta} 
+ \frac{1}{2}\pounds_{\vec{\boldsymbol n}} \mathcal{E}_{\beta\gamma}\right)\;,
\end{equation}
where $\pounds_{\vec{\boldsymbol n}} \mathcal{E}_{\beta\gamma}$ can in turn be computed with (\ref{eq:lied-weyl-electric}).
After this replacement is done one gets a rather involved equation which can be simplified if 
one does the following substitutions on it
\begin{equation}
\mathcal{E}_{\mu\nu} \mathcal{E}^{\mu\nu}=\frac{a}{16}\;,\quad
\mathcal{E}_{\mu}{}^{\alpha} \mathcal{E}^{\mu\nu} \mathcal{E}_{\nu\alpha}=-\frac{b}{64}\;,\quad
\mathcal{E}_{\mu\nu}D_\rho\mathcal{E}^{\mu\nu}=\frac{D_\rho a}{32}\;,\quad
\mathcal{E}_{\alpha}{}^{\gamma}\mathcal{E}^{\alpha\beta}D_\mu\mathcal{E}_{\beta\gamma}=-\frac{D_\mu b}{192}.
\end{equation}
After lengthy algebra one gets the intermediate expression
\begin{eqnarray}
&&\nabla_{\mu}\mathit{w} = \frac{\mathit{b} D_{\mu}\mathit{a} -  \mathit{a} \
D_{\mu}\mathit{b}}{2 \mathit{a}^2} +\nonumber\\ 
&&\frac{n_{\mu}}{\mathit{a}^2} \Bigl(K^{\beta\gamma} \bigl(-6 \mathit{a}^2 \mathcal{E}_{\beta\gamma} 
+ 48 \mathcal{E}_{\beta}{}^{\delta} (\mathit{b} \mathcal{E}_{\gamma\delta} + 6 \mathit{a} 
\mathcal{E}_{\gamma}{}^{\sigma} \mathcal{E}_{\delta\sigma})\bigr) + \mathit{a} \mathit{b} K^{\beta}{}_{\beta}  - \nonumber\\ 
&& 16\ \mbox{i}\ \mathcal{E}^{\beta\gamma} 
(\mathit{b} \varepsilon_{\gamma\delta\sigma} D^{\sigma}\mathcal{E}_{\beta}{}^{\delta} + 
6 \mathit{a} \varepsilon_{\delta\sigma\lambda} 
\mathcal{E}_{\beta}{}^{\delta} D^{\lambda}\mathcal{E}_{\gamma}{}^{\sigma})\Bigr).\nonumber\\
&&
\end{eqnarray}
The final result (\ref{eq:decomposew}) comes from here after doing the replacement
\begin{equation}
\mathcal{E}_{\alpha}{}^{\rho} \mathcal{E}_{\nu\rho} = \frac{\mathit{a}^2 h_{\alpha\nu}  - \
12 \mathit{b} \mathcal{E}_{\alpha\nu} - 12 \mathit{a} \mathfrak{a}_{\alpha\nu}}{48\mathit{a}}.
\end{equation}
\qed

\begin{proposition}
 The condition
\begin{equation}
\Xi {}_{\mu}{}_{\nu}= \frac{27}{2} w^{\frac{11}{3}} \xi {}_{\mu} \xi {}_{\nu}\;,
\label{eq:xitokilling}
 \end{equation}
is equivalent to
\begin{eqnarray}
&& \mathcal{E}_{\mu\nu}\left( (\omega^{\parallel})^2 +
(\omega^{\bot})_{\beta} (\omega^{\bot})^{\beta}\right)- 
2(\omega^{\bot})^{\beta}\left(
i\ \mathcal{E}_{(\mu}{}^{\delta} \varepsilon_{\nu)\beta\delta} \omega^{\parallel} 
-\mathcal{E}_{\beta(\mu}(\omega^{\bot})_{\nu)}\right) + \nonumber\\
&&\tfrac{1}{2} \mathit{w} (\omega^{\bot})_{\mu} (\omega^{\bot})_{\nu}+
h_{\mu\nu} \bigl(\tfrac{1}{2} \mathit{w} (\omega^{\parallel})^2 + \
(\omega^{\bot})^{\beta} (- \tfrac{1}{2} \mathit{w} \
(\omega^{\bot})_{\beta} + \mathcal{E}_{\beta\delta} (\omega^{\bot})^{\delta})\bigr) \
= \frac{27}{2} w^{\frac{11}{3}} Y_{\mu} Y_{\nu}\;,\nonumber\\
&& \label{eq:killing-split-n}
\end{eqnarray}
where
\begin{eqnarray}
&&Y^2 = \left(\mathit{w} (\omega^{\bot})_{\mu} (\omega^{\bot})^{\mu} +
2 \mathcal{E}_{\mu\nu} (\omega^{\bot})^{\mu} (\omega^{\bot})^{\nu}\right)w^{-\frac{11}{3}}\;,\label{eq:killing-candidate-1n}\\
&&Y_{\mu} = \frac{2 \mathcal{E}_{\mu\nu} \omega^{\parallel} \
(\omega^{\bot})^{\nu} - 2\ i\ \varepsilon_{\mu\beta\delta} \mathcal{E}_{\nu}{}^{\delta}
(\omega^{\bot})^{\nu} (\omega^{\bot})^{\beta} + 
\mathit{w} \omega^{\parallel} (\omega^{\bot})_{\mu}}{Y w^{\frac{11}{3}}}.
\label{eq:killing-candidate-2n}
\end{eqnarray}
\label{prop:xi-splitting} 
\end{proposition}

\proof To prove this one has to compute the orthogonal splitting of the intervening quantities
in eq. (\ref{eq:xitokilling}) and set the spatial parts to zero (see  (\ref{eq:spatial-tensor-zero})).
The orthogonal splitting of $\xi_\mu$ is
$$
\xi_\mu=Y n_\mu+Y_\mu\;,\label{eq:xi-split}
$$
and the orthogonal splitting of $\Xi_{\mu\nu}$ is found by looking at 
(\ref{eq:xiqw}) and computing the orthogonal splitting of $\mathcal{C}_{\mu\nu\alpha\beta}$, 
$w$, $\nabla_\mu w$ and $\mathcal{G}_{\mu\nu\alpha\beta}$. 
These splittings can be found in resp. eq. (\ref{eq:decompose-weyl-selfdual}), (\ref{eq:w-split}), 
Proposition \ref{prop:omega-decomposition} and eq. (\ref{decompose_eta}) 
(see the definition of $\mathcal{G}_{\mu\nu\alpha\beta}$ in eq. (\ref{eq:xiqw})). The result then follows after
lengthy but straightforward computations.\qed

\section{Construction of vacuum type D initial data}
\label{sec:typeD}
In the standard formulation of the Cauchy problem in general relativity 
one considers a 3-dimensional connected Riemannian manifold $(\Sigma,h_{ij})$
and an isometric embedding $\phi:\Sigma\longrightarrow\mathcal{M}$. The map $\phi$ is an
isometric embedding if $\phi^*g_{\mu\nu}=h_{ij}$ where
$\phi^*$ denotes the pull-back of tensor fields from $\mathcal{M}$ to
$\Sigma$.   We use small plain Latin letters $i,j,k,\dots$ for the abstract indices
of tensors on the manifold $\Sigma$.
The metric $h_{ij}$ defines a unique affine connection
$D_i$ without torsion (Levi-Civita connection) by means of the standard condition
\begin{equation}
D_jh_{ik}=0.
\end{equation}
The Riemann tensor of $D_i$ is denoted by $r_{ijkl}$ and from it we define its
Ricci tensor by $r_{ij}\equiv r^{l}_{\phantom{l}ilj}$ and its scalar
curvature $r\equiv r^i_{\phantom{i}i}$ (in $\Sigma$ indices are
raised and lowered with $h_{ij}$ and its inverse $h^{ij}$).
\begin{theorem} 
Let $(\Sigma,h_{ij})$ be a Riemannian manifold and suppose that
there exists a symmetric tensor field $K_{ij}$ on it which satisfies
the conditions (vacuum constraints)
\begin{eqnarray}
&& r+ K^2-K^{ij}K_{ij}=0, \label{Hamiltonian}\\
&& D^jK_{ij}-D_iK=0, \label{Momentum}
\end{eqnarray}
where $K\equiv K^{i}_{\phantom{i}i}$. Provided that $h_{ij}$ and $K_{ij}$ are 
smooth there exists an isometric embedding $\phi$ of
$\Sigma$ into a globally hyperbolic, vacuum solution
$(\mathcal{M},g_{\mu\nu})$ of the Einstein field equations. The set
$(\Sigma,h_{ij}, K_{ij})$ is then called a vacuum initial data
set and the spacetime $(\mathcal{M},g_{\mu\nu})$ is the data
development. Furthermore the spacelike hypersurface
$\phi(\Sigma)$ is a Cauchy hypersurface in $\mathcal{M}$.
\label{theorem:vacuum-data}
\end{theorem}
The previous theorem is true under more general differentiability assumptions 
for $h_{ij}$ and $K_{ij}$ (see Theorem 8.9 of \cite{CHOQUETBRUHATBOOK}).
Since $\phi(\Sigma)$ is a Cauchy hypersurface of $\mathcal{M}$ we shall often use the standard notation
$D(\Sigma)$ for $\mathcal{M}$ (the identification $\Sigma\leftrightarrow\phi(\Sigma)$ is then
implicitly understood).
The main object of this paper is the characterisation of general {\em type D initial data}. We 
give next the formal definition of this concept.
\begin{definition}
A vacuum initial data set $(\Sigma, h_{ij}, K_{ij})$ is called a type D initial data 
set if there exists an isometric embedding 
$\phi:\Sigma\rightarrow\mathcal{M}$ where $\mathcal{M}$ is a vacuum type D space-time.
\label{def:typeD-initial-data}
\end{definition}

To proceed further, we construct a foliation of $\mathcal{M}$ 
with a vector field $n^\mu$, $n^\mu n_\mu=-1$ defined on $\mathcal{M}$ which is orthogonal to the leaves and we 
denote by $\{\Sigma_t\}_{t\in I\subset\mathbb{R}}$ the family of leaves of
this foliation (we assume that $0\in I$). We choose the foliation in such a way that the leaf $\Sigma_0$ is related
to the Riemannian manifold $\Sigma$ introduced in Theorem \ref{theorem:vacuum-data} by 
$\phi(\Sigma)=\Sigma_0$. Under these conditions
the leaf $\Sigma_0$ is called the {\em initial data hypersurface} and we shall say that 
$n^\mu$ is a $\Sigma$-normal vector field in $\mathcal{M}$. 
The interest of introducing 
a foliation is that we can use the unit vector $n^\mu$ to perform the {\em orthogonal} splitting
of any tensorial quantity defined on $\mathcal{M}$ in the manner explained in section \ref{sec:ot-typeD}
and then relate the terms of this splitting
to tensors in $\Sigma$ by means of the embedding
$\phi$. For example one has $\phi^*h_{\mu\nu}=h_{ij}$. 
Other important examples are
\begin{equation}
\phi^*K_{\mu\nu}=K_{ij},\ 
\phi^*(D_\mu T_{\beta_1\dots\beta_q})=
D_i(\phi^*T_{\beta_1\dots\beta_q}).
\end{equation}
In these cases the pull-back is computed by just 
replacing the Greek indices by Latin ones. This generalises to any covariant tensor which is 
spatial with respect to a $\Sigma$-normal vector $n^\mu$. Also for any tensor ${\boldsymbol T}$ 
defined on $\mathcal{M}$ 
eq. (\ref{eq:tensor-general-ot}) entails
\begin{equation}
{\boldsymbol T}=0\Rightarrow
{\boldsymbol T}|_{\phi(\Sigma)}=0 \Longleftrightarrow
\phi^*({\boldsymbol T}^{(J)}_{(P)})=0  \;,\forall J\;,\quad
\label{eq:spatialp}
\end{equation}
\begin{theorem}[{\bf Vacuum type D initial data: necessary conditions}]
 Any vacuum type D initial data set $(\Sigma, h_{ij}, K_{ij})$ satisfies the conditions
 \begin{equation}
  \frac{a}{12} h_{ij} - \frac{b}{a}\mathcal{E}_{ij} - 4 \mathcal{E}_{i}{}^{k} \mathcal{E}_{jk}=0\;,
 \label{eq:typeD-necessary}
 \end{equation}
where
 \begin{eqnarray}
&& a \equiv 16 \mathcal{E}_{ij} \mathcal{E}^{ij}\;,\quad b \equiv -
64 \mathcal{E}_{i}{}^{k} \mathcal{E}^{ij} \mathcal{E}_{jk}.\label{eq:a-split}\\
&&  \mathcal{E}_{kl}=\frac{1}{2}(E_{kl}-\mbox{i}\; B_{kl})\;,\label{eq:e-split}\\
&& E_{ij}= r_{ij} + K K_{ij} - K_{ik} \updn{K}{k}{j}\;,\quad
B_{ij}= \updn{\epsilon}{kl}{(i} D_{|k} K_{l|j)}.\label{eq:eb-to-initial-data}
\end{eqnarray}
\label{theo:type-d-initial-data-necessary}
 \end{theorem}
\proof 
Use (\ref{eq:spatialp}) with ${\boldsymbol T}=\mathcal{D}_{\mu\nu\alpha\beta}$ in
combination with Proposition \ref{prop:d-tensor-split} and Theorem \ref{theorem:complex-type-D}
to get (\ref{eq:typeD-necessary})-(\ref{eq:e-split}). Expression (\ref{eq:eb-to-initial-data})
is the standard expression of the pull-back of the Weyl tensor electric and magnetic parts 
to the Riemannian manifold $\Sigma$ \cite{GARVALSCH, GARKERR}. \qed

In principle the conditions of Theorem \ref{theo:type-d-initial-data-necessary} are only 
necessary conditions which a type D initial data set must comply with but to have a complete characterisation 
of type D initial data we also need to find sufficient conditions. This task is carried out in the next 
subsection.

\subsection{Construction of necessary and sufficient Type D initial data conditions}

To find a necessary and sufficient set of conditions on a vacuum initial data set which guarantees that the data are 
type D initial 
data we follow a procedure already employed in \cite{GARKERR,GARVALSCH} which we summarise next. In the first step 
conditions are appended to the data shown in Theorem \ref{theo:type-d-initial-data-necessary} 
which guarantee the existence of a Killing vector in the data development. In the second
step we refine the conditions to ensure that the Killing vector coincides with the Killing vector defined by 
eq. (\ref{eq:killing-type-D}). These conditions turn out to be necessary and sufficient to guarantee that 
a subset of the data development is of Petrov type D. 

The first step shall be carried out with the aid of the notion of {\em Killing intial data} (KID). 
A Killing initial data set (KID) associated to a vacuum initial data 
$(\Sigma, h_{ij}, K_{ij})$ is a pair
$(\tilde{Y}, \tilde{Y}_i)$ consisting of a scalar $\tilde{Y}$ and a vector $\tilde{Y}_i$ defined on
$\Sigma$ satisfying the following system of partial differential
equations on $\Sigma$
\begin{subequations}
\begin{eqnarray}
&& D_{(i}\tilde{Y}_{j)}-\tilde{Y} K_{ij}=0, \label{kid-1}\\ 
&& D_iD_j\tilde{Y}-\mathcal{L}_{\tilde{Y}^l} K_{ij} = \tilde{Y}(r_{ij}+KK_{ij}-2K_{il}K^l_{\phantom{l}j})\label{kid-2}.
\end{eqnarray}
\end{subequations} 
The fundamental result about a KID set is the following Theorem whose proof can be found in  
\cite{CHRUSCIEL-BEIG-KID,COLL77,MONCRIEF-KID} (the formulation is taken from \cite{GARKERR})
\begin{theorem}
The necessary and sufficient condition for there to exist a Killing vector $\xi^\mu$ in 
the data development
of a vacuum initial data set $(\Sigma, h_{ij}, K_{ij})$ is that a pair $(\tilde{Y}, \tilde{Y}_j)$
fulfills eqs. (\ref{kid-1})-(\ref{kid-2}). The orthogonal splitting of $\xi^\mu$ 
with respect to any $\Sigma$-normal unit timelike vector field $n^\mu$ is 
\begin{equation}
 \xi_\mu= Y n_\mu+ Y_\mu\;,\quad Y\equiv -(n_\nu\xi^\nu)\;,\quad 
 Y_\mu \equiv h_\mu{}^\nu\xi_\nu\;,\quad 
 \phi^*(Y)=\tilde{Y}\;,\quad \phi^*Y_\mu=\tilde{Y}_j.
\label{eq:killing-splitting}
\end{equation} 
\label{thm:kid}
\end{theorem}
The previous Theorem has been formulated in the literature implicitly assuming that
$(\tilde{Y}, \tilde{Y}_j)$, $\xi^\mu$ are real but since 
(\ref{kid-1})-(\ref{kid-2}) are linear in $(\tilde{Y}, \tilde{Y}_j)$, the result can be formulated assuming that the pair 
$(\tilde{Y}, \tilde{Y}_j)$
is formed by complex valued tensor fields. In this case the Killing field $\xi^\mu$ will in general be complex.

To accomplish the second step of our procedure we need a lemma.

\begin{lemma}
If $\vec{\boldsymbol n}$ is a $\Sigma$-normal unit vector field on $\mathcal{M}$  
and $\vec{\boldsymbol \xi}$ is a vector field on $\mathcal{M}$ such that 
\begin{equation}
 \xi_\mu= Y n_\mu+ Y_\mu\;,\quad 
\tilde{Y}\equiv \phi^*(Y)\;,\quad \tilde{Y}_j\equiv \phi^*Y_\mu\;,
\label{eq:xi-development}
\end{equation}

then
one has the following equivalence 
\begin{eqnarray}
&&\left.\left(\Xi {}_{\mu}{}_{\nu}
- \frac{27}{2} w^{\frac{11}{3}} \xi {}_{\mu} \xi {}_{\nu}\right)\right|_{\phi(\Sigma)}=0\Longleftrightarrow\\
&&\mathcal{E}_{pj}\left( (\omega^{\parallel})^2 +
(\omega^{\bot})_{l} (\omega^{\bot})^{l}\right)- 
2(\omega^{\bot})^{l}\left(
i\ \mathcal{E}_{(p}{}^{k} \varepsilon_{j)lk} \omega^{\parallel} 
-\mathcal{E}_{l(p}(\omega^{\bot})_{j)}\right) + \nonumber\\
&&\tfrac{1}{2} \mathit{w} (\omega^{\bot})_{p} (\omega^{\bot})_{j}+
h_{pj} \bigl(\tfrac{1}{2} \mathit{w} (\omega^{\parallel})^2 + \
(\omega^{\bot})^{l} (- \tfrac{1}{2} \mathit{w} \
(\omega^{\bot})_{l} + \mathcal{E}_{lk} (\omega^{\bot})^{k})\bigr) \
= \frac{27}{2} w^{\frac{11}{3}}\tilde{Y}_{p} \tilde{Y}_{j}\;,\label{eq:killing-split}\\
&&\tilde{Y}^2 = \left(\mathit{w} (\omega^{\bot})_{j} (\omega^{\bot})^{j} +
2 \mathcal{E}_{jk} (\omega^{\bot})^{j} (\omega^{\bot})^{k}\right)w^{-\frac{11}{3}}\;,\label{eq:killing-candidate-1}\\
&&\tilde{Y}_{j} = \frac{2 \mathcal{E}_{jk} \omega^{\parallel} \
(\omega^{\bot})^{k} - 2\ i\ \varepsilon_{jkl} \mathcal{E}_{p}{}^{l}
(\omega^{\bot})^{p} (\omega^{\bot})^{k} + 
\mathit{w} \omega^{\parallel} (\omega^{\bot})_{j}}{\bar{Y} w^{\frac{11}{3}}}\;,\label{eq:killing-candidate-2}
\end{eqnarray}
where
\begin{eqnarray}
&&(\omega^\bot)_j\equiv 
\frac{b D_ {j}\mathit{a} -  \mathit{a} D_ {j}\mathit{b}}{2a^2}=D_j w\;,\\
&&\omega^\parallel\equiv 
\frac{6 K^{jk}}{\mathit{a}^3} \bigl(\mathit{b}^2 \mathcal{E}_{jk} + \mathit{a} \
(\mathit{b} \mathfrak{a}_{jk} - 12 \mathit{a} \mathcal{E}_{j}{}^{l}
\mathfrak{a}_{kl})\bigr) -w K^{j}{}_{j} + \mbox{i}\ 
\frac{16(\mathit{b} \mathcal{E}^{jk} + 3 \mathit{a} \mathfrak{a}^{jk})}{2 \mathit{a}^2}
\varepsilon_{k p l} D^{l}\mathcal{E}_{j}{}^{p}\;,\nonumber\\
&&\mathfrak{a}_{lj}\equiv\frac{a}{12} h_{lj} - \frac{b}{a}\mathcal{E}_{lj} - 4 \mathcal{E}_{l}{}^{k} \mathcal{E}_{jk}\;,\quad
a \equiv 16 \mathcal{E}_{lj} \mathcal{E}^{lj}\;,\quad 
b \equiv -64 \mathcal{E}_{l}{}^{k} \mathcal{E}^{lj} \mathcal{E}_{jk}\;,\quad w\equiv -\frac{b}{2a}\;,
\end{eqnarray}
and  $\mathcal{E}_{kl}$ is computed using (\ref{eq:e-split})-(\ref{eq:eb-to-initial-data}).
\label{lemma:omega}
\end{lemma}
\proof We compute the orthogonal splitting of the tensor expression $\Xi_{\mu\nu}-27/2 w^{11/3}\xi_\mu\xi_\nu$ and 
then use (\ref{eq:spatialp}) to find the result. The orthogonal splitting of this tensor expression is the content
of Proposition \ref{prop:xi-splitting} so we only need to compute the pull-back of the intervening quantities to 
$\Sigma$. In doing so one needs to use the results of Proposition \ref{prop:omega-decomposition} and recall our 
convention of replacing Greek by Latin abstract indices when computing pull-backs of spatial quantities 
with respect to a $\Sigma$-normal unit vector.\qed

\begin{theorem}
Suppose that $\tilde{Y}$, $\tilde{a}$, $\tilde{w}$ defined below vanish nowhere on $\Sigma$. Then
the data development of a vacuum initial data set $(\Sigma, h_{ij}, K_{ij})$ is of Petrov type D if, 
and only if, the following conditions hold
\begin{eqnarray}
&&\frac{\tilde{a}}{12} h_{ij} - \frac{\tilde{b}}{\tilde{a}}\mathcal{E}_{ij} - 4 \mathcal{E}_{i}{}^{k} \mathcal{E}_{jk}=0\;,
\label{eq:mathfraka-condition}\\
&&\mathcal{E}_{pj}\left( \Omega^2 +\Omega_{l} \Omega^{l}\right)- 
2\Omega^{l}\left(\mbox{i}\ \mathcal{E}_{(p}{}^{k} \varepsilon_{j)lk} \Omega 
-\mathcal{E}_{l(p}\Omega_{j)}\right) + \nonumber\\
&&\tfrac{1}{2} \tilde{w} \Omega_{p} \Omega_{j}+h_{pj} 
\bigl(\tfrac{1}{2} \tilde{w} \Omega^2 + \Omega^{l} (- \tfrac{1}{2} \tilde{w} 
\Omega_{l} + \mathcal{E}_{lk} \Omega^{k})\bigr) \
= \frac{27}{2} \tilde{w}^{\frac{11}{3}} \tilde{Y}_{p} \tilde{Y}_{j}\;,\label{eq:type-d-sufficient}\\
&& D_{(i}\tilde{Y}_{j)}-\tilde{Y} K_{ij}=0, \label{kid-1-theorem}\\ 
&& D_iD_j\tilde{Y}-\mathcal{L}_{\tilde{Y}^l} K_{ij} = \tilde{Y}(r_{ij}+KK_{ij}-2K_{il}K^l_{\phantom{l}j})\label{kid-2-theorem}\;,
\end{eqnarray}
where
\begin{eqnarray}
&&
\tilde{a} \equiv 16 \mathcal{E}_{ij} \mathcal{E}^{ij}\;,\quad 
\tilde{b} \equiv -64 \mathcal{E}_{i}{}^{k} \mathcal{E}^{ij} \mathcal{E}_{jk}\;,\quad
\tilde{w}\equiv -\frac{\tilde{b}}{2\tilde{a}}\;,\\
&&
\Omega_j\equiv D_j\tilde{w}\;,\quad \Omega\equiv 
K^{jk}\mathcal{E}_{jk} -\tilde{w} K^{j}{}_{j} - 16\mbox{i}\ 
\frac{\tilde{w}}{\tilde{a}}\mathcal{E}^{jk}
\varepsilon_{k p l} D^{l}\mathcal{E}_{j}{}^{p}\;,\nonumber\\
&&\tilde{Y} \equiv \left(\tilde{w} \Omega_{j}\Omega^{j} +
2 \mathcal{E}_{jk} \Omega^{j}\Omega^{k}\right)^{\frac{1}{2}}\tilde{w}^{-\frac{11}{6}}\;,\label{eq:killing-candidate-1-s}\\
&&\tilde{Y}_{j} \equiv \frac{\Omega(2 \mathcal{E}_{jk}\Omega^{k}+\tilde{w}\Omega_{j}) 
- 2\mbox{i}\ \varepsilon_{jkl} \mathcal{E}_{p}{}^{l}
\Omega^{p} \Omega^{k}}{\tilde{Y} \tilde{w}^{\frac{11}{3}}}\;,
\end{eqnarray}
and  $\mathcal{E}_{kl}$ is computed using (\ref{eq:e-split})-(\ref{eq:eb-to-initial-data}).
\label{thm:type-d-initialdata-sufficient}
\end{theorem}

\proof 
We first start proving the necessity of the conditions.
Theorem \ref{theo:type-d-initial-data-necessary} proves that (\ref{eq:mathfraka-condition})
is necessary (compare eqs. (\ref{eq:mathfraka-condition}) and (\ref{eq:typeD-necessary})). 
Theorem \ref{theorem:complex-killing} implies that the conditions of Lemma \ref{lemma:omega} 
are also necessary but in this particular case they can be simplified as follows:
given the fact that $\mathcal{E}_{kl}$ is symmetric and traceless we can find an orthonormal frame 
on $\Sigma$ in which one has 
\begin{equation}
h_{kl}=\mbox{diag}(1,\;1,\;1),\;\quad
\mathcal{E}_{kl}=\mbox{diag}(\lambda_1,\;\lambda_2,\;-\lambda_1-\lambda_2).
\end{equation}
Using this information in (\ref{eq:mathfraka-condition}) we get
\begin{equation}
\frac{\tilde{a}}{12}  -  \frac{\tilde{b} \lambda_1}{\tilde{a}} - 4 \lambda_1^2 = 0\;,\quad
\frac{\tilde{a}}{12}  -  \frac{\tilde{b} \lambda_2}{\tilde{a}} - 4 \lambda_2^2 = 0\;,\quad
\frac{\tilde{a}}{12}  + \frac{\tilde{b} \lambda_1}{\tilde{a}} - 
4 \lambda_1^2 + \frac{\tilde{b} \lambda_2}{\tilde{a}} - 8 \lambda_1 \lambda_2 - 4 \lambda_2^2 = 0.
\end{equation}
Combining these relations yields
\begin{equation}
 6 \tilde{b}^2=\tilde{a}^3.
\label{eq:b2a3}
\end{equation}
If we use this equation and the properties $a=\tilde{a}$, $b=\tilde{b}$, $\mathfrak{a}=0$, 
$w=\tilde{w}$ we deduce that (\ref{eq:killing-split})-(\ref{eq:killing-candidate-2}) 
in Lemma \ref{lemma:omega} become (\ref{eq:type-d-sufficient}) with the replacements
\begin{equation}
(\omega^\bot)_j=\Omega_j\;,\quad
\omega^\parallel=\Omega\;,\quad
\bar{Y}=\tilde{Y}\;,\quad
\bar{Y}_j=\tilde{Y}_j.
\label{eq:replacements}
\end{equation}
Also the conditions (\ref{eq:xi-development}), 
combined with Theorem \ref{thm:kid} yield (\ref{kid-1-theorem})-(\ref{kid-2-theorem}).

To show sufficiency we first note that if (\ref{eq:mathfraka-condition}) holds 
then (\ref{eq:b2a3}) can be deduced by means of the same reasoning as above
so using again the replacements (\ref{eq:replacements}) we conclude
that the conditions of Lemma \ref{lemma:omega} are still true. Thus 
\begin{equation}
\left.\left(\Xi {}_{\mu}{}_{\nu}
- \frac{27}{2} w^{\frac{11}{3}} \xi {}_{\mu} \xi {}_{\nu}\right)\right|_{\phi(\Sigma)}=0.
\label{eq:xi-initial-data}
\end{equation}
where, due to the assumed conditions (\ref{kid-1-theorem})-(\ref{kid-2-theorem}) 
and Theorem \ref{thm:kid}, $\xi^\mu$ is a Killing vector field in the data development $D(\Sigma)$.
Also if we combine (\ref{eq:spatialp}) and 
(\ref{eq:mathfraka-condition}) we deduce 
\begin{equation}
 \mathcal{D}_{\mu\nu\rho\sigma}|_{\phi(\Sigma)}=0.
\label{eq:d-initial-data}
\end{equation}
The fact that $\vec{\boldsymbol\xi}$ is a Killing vector implies that the Lie  
derivative defined from it vanishes when acting on a Weyl concomitant. Thus   
\begin{equation}
 \pounds_{\vec{\boldsymbol\xi}}\mathcal{D}_{\mu\nu\rho\sigma}=0\;,\quad
\pounds_{\vec{\boldsymbol\xi}}\left(\Xi {}_{\mu}{}_{\nu}
-\frac{27}{2} w^{\frac{11}{3}} \xi {}_{\mu} \xi {}_{\nu}\right)=0.
\label{eq:d-xi-evolution}
\end{equation}
These results enable us to regard (\ref{eq:xi-initial-data})-(\ref{eq:d-xi-evolution}) as a 
first order linear evolution system in the variables $\mathcal{D}_{\mu\nu\rho\sigma}$, 
$\Xi {}_{\mu}{}_{\nu}
-\frac{27}{2} w^{\frac{11}{3}} \xi {}_{\mu} \xi {}_{\nu}$. 
The 
data of the system are trivial and non-characteristic given that by assumption $\tilde{Y}\neq 0$
(the characteristic points of the system (\ref{eq:xi-initial-data})-(\ref{eq:d-xi-evolution})
are those in which $\vec{\boldsymbol\xi}$ is tangent to $\phi(\Sigma)$).
Hence we conclude that there is an open subset $\mathcal{U}\subset D(\Sigma)$ containing $\Sigma$
where one has 
\begin{equation}
 \mathcal{D}_{\mu\nu\rho\sigma}=0\;,\quad 
 \Xi{}_{\mu}{}_{\nu}=\frac{27}{2} w^{\frac{11}{3}} \xi {}_{\mu} \xi {}_{\nu}\;,\quad
 \nabla_{\mu}\xi_\nu + \nabla_{\nu}\xi_\mu = 0.
\label{eq:type-D-development}
\end{equation}
In fact Theorem \ref{theorem:complex-killing} tells us that the two last conditions are 
{\em redundant} if the first one holds. Thus we conclude from Theorem \ref{theorem:complex-type-D} 
that the Weyl tensor is of Type D on $\mathcal{U}$ and indeed it may correspond to {\em any} type D solution. 
\qed

\section{Applications}
\label{sec:applications}
In this section we show that our results can be used to find a new characterisation of the Kerr solution 
which can in turn be used to construct Kerr initial data
sets. 

\begin{theorem}
Under the hypotheses of Theorem \ref{theorem:complex-killing} a space-time is locally isometric to the Kerr 
solution with non-vanishing mass (non-trivial Kerr solution) if and only if the following additional conditions hold
\begin{eqnarray}
&& \xi_{[\mu}\bar\xi_{\nu]}=0\;,
\label{eq:killing-property-kerr}\\
&&\mbox{\em Im}(Z^{3}\bar w^8)=0\;,\quad Z\equiv \nabla_\rho w \nabla^\rho w\;,
\label{eq:nutzero}\\
&&
\frac{\mbox{\em Re}(Z^{3}\bar w^8)}{\big(18 \mbox{\em Re}\big(w^3\bar Z\big)-|Z|^2\big)^3}<0,\;
(\mbox{if}\ 18 \mbox{\em Re}\big(w^3\bar Z\big)-|Z|^2\neq 0),\;
\label{eq:epsilongeq0}\\
&&
\mbox{\em Re}(Z^{3}\bar w^8)=0\Longleftrightarrow \xi_\mu\bar\xi^\mu=0,\;
(\mbox{if}\
18 \mbox{\em Re}\big(w^3\bar Z\big)-|Z|^2= 0).
\label{eq:epsilongeq00}
\end{eqnarray}
where $\xi_\mu$ is defined by (\ref{eq:killing-type-D}).
\label{theo:kerr-local}
\end{theorem}
\proof
The space-time is locally isometric to the Kerr-NUT solution
if and only if the hypotheses of Theorem \ref{theorem:complex-killing} hold and $\xi_{[\mu}\bar\xi_{\nu]}=0$  
\cite{INVARIANTSYMMETRIESD}. The Kerr-NUT solution can be given in local coordinates
by (see \cite{EXACTSOLUTIONS}, eq. (21.16)) 
\begin{equation}
ds^2= \frac{(x^2 + y^2)}{X(x)}dx^2 + \frac{(x^2 + y^2)}{Y(y)}dy^2
 + \frac{1}{x^2 + y^2}\big(X(x)(dt -  y^2 dz)^2 - Y(y)(dt + x^2 dz)^2\big)\;,
\label{eq:kerr-metric}
\end{equation}
where 
\begin{equation}
 Y(y)\equiv\epsilon y^2 - 2 \mu y + \gamma\;,\quad 
 X(x) \equiv -\epsilon x^2 + 2\lambda x +\gamma\;,
\end{equation}
and $\epsilon$, $\gamma$, $\lambda$, $\mu$ are constants. If $\lambda=0$ and $\epsilon>0$ then the metric reduces to the 
Kerr solution whose mass and angular Momentum parameters are respectively
\begin{equation}
M=\frac{\mu}{\epsilon^{\frac{3}{2}}}\;,\quad A=\frac{2\sqrt{|\gamma|}}{\epsilon}. 
\end{equation}
Hence the mass is different from zero if $\mu\neq 0$. Note that if $\mu=0$ the space-time is just the flat Minkowski
solution which does not fulfill the hypotheses of Theorem \ref{theorem:complex-killing} so we assume henceforth that 
$\mu\neq 0$. An explicit computation using the coordinates of (\ref{eq:kerr-metric}) shows that
\begin{eqnarray}
&&w = \frac{\lambda + \mbox{i}\ \mu}{(x + \mbox{i}\ y)^3}\;,\quad
\nabla_{\rho} w \nabla^{\rho} w = \frac{9 (\lambda + \mbox{i}\ \mu)^2 \bigl(2 \lambda x + 2 \mu y -\epsilon (x^2 + y^2)\bigr)}
{(x - \mbox{i}\ y) (x + \mbox{i}\ y)^9}\;,\label{eq:kerr-nut-1}\\
&&\boldsymbol{\xi}=\frac{\pm\mbox{i}\; z}{(\lambda+\mbox{i}\mu)^{\frac{1}{3}}}\frac{\partial}{\partial t}\;,\quad
\xi_\mu\bar\xi^\mu=\frac{2\lambda  x+2 \mu  y-\epsilon\left(x^2+y^2\right)}{(\lambda^2 +\mu^2)^{\frac{1}{3}}\left(x^2+y^2\right)}
\;,\label{eq:kerr-nut-2}\\
&& Z^{3}\bar w^8=\frac{729 (\lambda^2 -\mu^2 -2\mbox{i}\lambda\mu) (\lambda^2 +\mu^2 )^6
\left(2\lambda  x+2 \mu  y-\epsilon\left(x^2+y^2\right)\right)^3}{(x^2+y^2)^{27}}\;,\label{eq:kerr-nut-3}\\
&&-|Z|^2+18 \mbox{\em Re}\big(w^3\bar Z\big)=-\frac{81 \epsilon  \left(\lambda ^2+\mu ^2\right)^2
   \left(\epsilon  \left(x^2+y^2\right)-2 \lambda  x-2
   \mu  y\right)}{\left(x^2+y^2\right)^9}\;,
\label{eq:wdwtoxy}
\end{eqnarray}
where $z$ is an unimodular complex constant. It is straightforward to check using these expressions that if $\lambda=0$, $\epsilon>0$ and $\mu\neq 0$ 
then (\ref{eq:nutzero})-(\ref{eq:epsilongeq00}) hold. Supposse now that (\ref{eq:nutzero})-(\ref{eq:epsilongeq00}) are true. 
If (\ref{eq:nutzero}) holds then from (\ref{eq:kerr-nut-3}) we deduce that $\lambda\mu=0$ which in our case implies $\lambda=0$. At those points where 
$-|Z|^2+18 \mbox{Re}\big(w^3\bar Z\big)\neq 0$ one can deduce from (\ref{eq:wdwtoxy}) that $\mu\neq 0$, $\epsilon\neq 0$ and
from (\ref{eq:kerr-nut-1})-(\ref{eq:wdwtoxy}) the relation
\begin{equation}
\frac{\mbox{Re}(Z^3\bar w^8)}{\left(18\mbox{Re}(w^3\bar Z)-|Z|^2\right)^3}=\frac{-\mu^2}{558 \epsilon^3}.\label{eq:lambdamu}
\end{equation}
Condition (\ref{eq:epsilongeq0}) entails then $\epsilon>0$. The conclusion is that $\lambda=0$, $\epsilon>0$ and $\mu\neq 0$
which as stated above corresponds to the (non-trivial) Kerr solution. 
If $-|Z|^2+18 \mbox{Re}\big(w^3\bar Z\big)= 0$ then (\ref{eq:kerr-nut-2}) tells us that this condition 
holds only in the set of points fulfilling the condition 
$$
2\lambda  x+2 \mu  y-\epsilon\left(x^2+y^2\right)=0.
$$ 
Since this is a co-dimension 1
subset of our manifold, the conditions $\lambda=0$, $\epsilon>0$ just proven when $-|Z|^2+18 \mbox{Re}\big(w^3\bar Z\big)\neq 0$ must also hold 
in this subset as the space-time is smooth and $\lambda$, $\epsilon$ are constants.
\qed

Theorem \ref{theo:kerr-local} is similar to Theorem 2 of \cite{FERSAEZKERR} and in fact 
the computations carried out in the proof of Theorem \ref{theo:kerr-local} follow a pattern similar 
to those of Theorem 2 of \cite{FERSAEZKERR} but using the complex formalism rather than 
the real formalism used in \cite{FERSAEZKERR}. For us the relevance of 
Theorem \ref{theo:kerr-local} is that it enables us to single out the case 
of Kerr initial data in the conditions specified 
by Theorem \ref{thm:type-d-initialdata-sufficient}.
\begin{theorem}
Under the conditions and definitions of Theorem \ref{thm:type-d-initialdata-sufficient} the data 
development of a vacuum initial data set is a subset of the Kerr space-time if and only if the following 
additional conditions hold at those points of $\Sigma$ in which $18 \mbox{\em Re}\big(\tilde{w}^3\overline{\tilde{Z}}\big)-|\tilde{Z}|^2\neq 0$
\begin{equation}
\mbox{\em Im}(\tilde{Y}\tilde{Y}_j)=0\;,\quad 
\mbox{\em Im}\big(\tilde{Z}^3(\overline{\tilde{w}})^8\big)=0\;,\quad
\frac{\mbox{\em Re}\big(\tilde{Z}^3(\overline{\tilde{w}})^8\big)}
{\bigg(18 \mbox{\em Re}\big(\tilde{w}^3\overline{\tilde{Z}}\big)-|\tilde{Z}|^2\bigg)^3} < 0\;,
\label{eq:kerr-initial-data}
\end{equation}
where $\tilde{Z}$ is defined by
\begin{equation}
\tilde{Z}\equiv \Omega^j\Omega_j-\Omega^2.
\end{equation}
Also if the subset of $\Sigma$ where $18 \mbox{\em Re}\big(\tilde{w}^3\overline{\tilde{Z}}\big)-|\tilde{Z}|^2$ vanish is non empty then one has that
on that subset
\begin{equation}
 \mbox{\em Re}\big(\tilde{Z}^3(\overline{\tilde{w}})^8\big)=0\;,\quad 
-\tilde{Y}^2+\tilde{Y}_j\tilde{Y}^j=0.
\end{equation}
\label{thm:kerr-data}
\end{theorem}
\proof If the conditions of Theorem \ref{thm:type-d-initialdata-sufficient} hold then
we know that an open subset of the data de\-ve\-lopment is of type D and the 
Killing vector $\xi$ generated by the Killing candidate ($\tilde{Y}$, $\tilde{Y}^j$)
has the property (\ref{eq:commute-killing}) (see \ref{eq:type-D-development}). This property implies
\begin{equation}
\pounds_{\vec{\boldsymbol\xi}}H_{\mu\nu}=0\;,\quad H_{\mu\nu}\equiv \xi_{[\mu}\bar\xi_{\mu]}. 
\label{eq:h-evolution}
\end{equation}
In addition, the first condition of (\ref{eq:kerr-initial-data}) entails
\begin{equation}
 H_{\mu\nu}|_{\Sigma}=0\;,
\label{eq:h-data}
\end{equation}
as one can explicitly check by performing the orthogonal splitting of $H_{\mu\nu}$ and using 
(\ref{eq:xi-development}). Combining (\ref{eq:h-evolution}) and (\ref{eq:h-data}) we conclude that 
in an open subset of the data development $H_{\mu\nu}=0$ and thus (\ref{eq:killing-property-kerr}) holds 
in that open set. Showing that the other conditions of Theorem \ref{theo:kerr-local} hold in
open subsets of the data development proceeds along the lines of Theorem \ref{thm:type-d-initialdata-sufficient}
and so we skip the details. The proof of the necessity of the conditions is also similar to that of Theorem 
\ref{thm:type-d-initialdata-sufficient}.\qed

\begin{remark}
\em 
For a Kerr initial data set,
the set of points where $18 \mbox{\em Re}\big(\tilde{w}^3\overline{\tilde{Z}}\big)-|\tilde{Z}|^2$ vanish, if non-empty, cannot be 
an open subset of $\Sigma$ because this set would correspond to the intersection of the initial data hypersurface (spacelike) with the ergosphere
which is timelike (see eqs. (\ref{eq:kerr-nut-2}) and (\ref{eq:wdwtoxy})).
\end{remark}

Kerr initial data have been already constructed in a number of places of the literature 
\cite{GARKERR,GAR-VAL-KERR} but either the expressions contain additional variables other than the initial data
quantities or they are cumbersome. The characterisation of Kerr initial data presented in Theorem \ref{thm:kerr-data}
is algorithmic in the initial data quantities $h_{jk}$, $K_{jk}$ and it simplifies the expressions obtained in 
\cite{GARKERR}.

\section{Outlook}
We have constructed Type D initial data in Theorem \ref{thm:type-d-initialdata-sufficient} and shown how 
these results can be used to construct Kerr initial data in Theorem \ref{thm:kerr-data}. In both cases 
the conditions only involve the standard quantities characterising a vacuum initial data set on a Riemannian
manifold $\Sigma$ (the Riemannian metric $h_{ij}$ and a symmetric tensor $K_{ij}$ playing the role of the second 
fundamental form). In this sense we say that our characterisation of vacuum type D initial data is {\em algorithmic}
because given a vacuum initial data set fulfilling the regularity conditions
of Theorem  \ref{thm:type-d-initialdata-sufficient}
we can check in an algorithmic fashion whether the data development is 
of type D or not. The characterisation we have found can be thought of as an overdetermined system of partial differential equations 
on a Riemannian manifold for a symmetric tensor $K_{ij}$ fulfilling the vacuum constraints 
(\ref{Hamiltonian})-(\ref{Momentum}). Choosing suitable gauge conditions could lead to a simplification of the system 
and questions such as the existence of solutions for the system could be analysed. 

\section*{Acknowledgements}
We thank professors Joan J. Ferrando and Jos\'e M. M. Senovilla for reading the manuscript and useful comments.  
We also thank Maria Okounkova for spotting a sign error in a formula in a previous version of the manuscript
and Swetha Bhagwat and Maria Okounkova for useful comments.
Supported by the project FIS2014-57956-P of Spanish ``Ministerio de Econom\'{\i}a y Competitividad''
and PTDC/MAT-ANA/1275/2014 of Portuguese ``Funda\c{c}\~{a}o para a Ci\^{e}ncia e a Tecnologia''.


\begin{thebibliography}{10}

\bibitem{LINEARISEDD}
S. Aksteiner and L. Andersson. 
\emph{{Linearized gravity and gauge conditions}},
 Classical and Quantum Gravity \textbf{28} (2011), 065001.

\bibitem{TYPEDLINEAR}
S. Aksteiner, L. Andersson and T. B{\"a}ckdahl, \emph{{On the
  structure of linearized gravity on vacuum spacetimes of Petrov type D}},
  \url{http://arxiv.org/abs/1601.06084}.

\bibitem{CHRUSCIEL-BEIG-KID}
R. Beig and P.~T. Chru{\'s}ciel, \emph{{Killing initial data}}, Classical
  and Quantum Gravity \textbf{14} (1997), A83--A92.

\bibitem{CHOQUETBRUHATBOOK}
Y. Choquet-Bruhat, \emph{{General Relativity and the Einstein Equations}},
  {Oxford Mathematical Monographs}, Oxford University Press, (2009).

\bibitem{COLL77}
B. Coll, \emph{{On the evolution equations for Killing fields}},
  Journal of Mathematical Physics \textbf{18} (1977), 1918--1922.

\bibitem{NATARIOCOSTAANALOGY}
L.~F.~ Costa and J.~ Nat{\'a}rio, \emph{{Gravito-electromagnetic
  analogies}}, General Relativity and Gravitation \textbf{46} (2014), 1--57.

\bibitem{SCHWARZSCHILDLINEAR}
M.~Dafermos, G.~Holzegel, and I.~Rodnianski, \emph{{The linear
  stability of the Schwarzschild solution to gravitational perturbations}},
  \url{http://arxiv.org/abs/1601.06467}.

\bibitem{EDGARHOGLUNDDDI}
S.~B. ~Edgar and A.~H{\"o}glund, \emph{{Dimensionally dependent tensor
  identities by double antisymmetrization}}, Journal of Mathematical Physics
  \textbf{43} (2002), 659--677.

\bibitem{ELLIS-COSMOLOGY}
G.~F.~R. Ellis, \emph{{Relativistic Cosmology}}, {Proceedings of the
  international school of physics "Enrico Fermi"} (B.~K. Sachs, ed.), {General
  Relativity and Cosmology}, Academic Press (1971), pp.~105--182.

\bibitem{FERRANDOCOVARIANTPETROVTYPE}
J.~J.~Ferrando, J.~A.~ Morales, and J.~A.~S{\'a}ez,
  \emph{{Covariant determination of the Weyl tensor geometry}}, Classical and
  Quantum Gravity \textbf{18} (2001), 4939--4959.

\bibitem{INVARIANTSYMMETRIESD}
J.~J.~ Ferrando and J.~A.~ S{\'a}ez, \emph{{On the invariant symmetries
  of the D-metrics}}, Journal of Mathematical Physics \textbf{48} (2007), 102504.

\bibitem{FERRSAEZTYPED}
J.~J.~Ferrando and J.~A. S{\'a}ez, \emph{{Type D vacuum solutions: a new
  intrinsic approach}}, General Relativity and Gravitation \textbf{46} (2014), 1--19.

\bibitem{FERSAEZKERR}
J.~J. Ferrando and J.~A. S{\'a}ez, \emph{{An intrinsic characterization of
  the {K}err metric}}, Class. Quantum Grav. \textbf{26} (2009), 075013, 13.

\bibitem{SE-DYNAMICALAWS}
A.~Garc{\'i}a-Parrado, \emph{{Dynamical laws of superenergy in general
  relativity}}, Class. Quantum. Grav. \textbf{25} (2008), 015006, 26.

\bibitem{GARVALSCH}
A.~ Garc{\'i}a-Parrado and J.~A. {Valiente Kroon}, \emph{{Initial data
  sets for the {S}chwarzschild spacetime}}, Phys. Rev. D \textbf{75} (2007),
  024027, 14.

\bibitem{GAR-VAL-KERR}
\bysame, \emph{{{K}err initial data}}, Classical Quantum Gravity \textbf{25}
  (2008), 205018, 20.

\bibitem{GARKILLINGSPINOR}
\bysame \emph{{Killing spinor initial data sets}}, Journal of 
Geometry and Physics \textbf{58}, (2008) 1186--1202.

  
\bibitem{GARKERR}
A.~ {Garc{\'i}a-Parrado G{\'o}mez-Lobo}, \emph{{Local non-negative initial
  data scalar characterization of the Kerr solution}}, Phys. Rev. D \textbf{92}
(2015), 124053, 13.

\bibitem{KLAINERMANAXIALKERR}
A.~D. Ionescu and S.~Klainerman, \emph{{On the Global Stability of
  the Wave-map Equation in Kerr Spaces with Small Angular Momentum}}, Annals of
  PDE \textbf{1} (2015), 1--78.

\bibitem{KERR-METRIC}
R.~P. Kerr, \emph{{Gravitational field of a spinning mass as an example of an
  algebraically special metric}}, Phys. Rev. Lett. \textbf{11} (1963), 237--238.

\bibitem{TYPE-D-KINNERSLEY}
W.~Kinnersley, \emph{{Type {D} {V}acuum {M}etrics}}, J. Math. Phys. \textbf{10}
  (1969), 1195--1203.

\bibitem{XACT}
J.~M. Mart{\'i}n-Garc{\'i}a, \emph{{x{A}ct: efficient tensor computer
  algebra}}, \url{http://www.xact.es}.

\bibitem{MONCRIEF-KID}
V.~Moncrief, \emph{{Spacetime symmetries and linearization stability of
  the Einstein equations.I}}, Journal of Mathematical Physics \textbf{16}
  (1975), 493--497.

\bibitem{EXACTSOLUTIONS}
H.~Stephani, D.~Kramer, M.~MacCallum, C.~Hoenselaers, and E.~Herlt,
  \emph{{Exact solutions of {E}instein's field equations}}, second ed.,
  {Cambridge Monographs on Mathematical Physics}, Cambridge University Press,
  Cambridge, 2003.

\end{thebibliography}

\providecommand{\bysame}{\leavevmode\hbox to3em{\hrulefill}\thinspace}
\providecommand{\MR}{\relax\ifhmode\unskip\space\fi MR }
\providecommand{\MRhref}[2]{%
  \href{http://www.ams.org/mathscinet-getitem?mr=#1}{#2}
}
\providecommand{\href}[2]{#2}

\end{document}